\documentclass[conference,a4paper]{APSIPA2025}
\usepackage{amsmath}
\usepackage{graphicx}
\usepackage{multirow}
\usepackage{threeparttable}
\usepackage{framed}
\usepackage{amsmath}
\usepackage{amssymb}
\usepackage{latexsym}
\usepackage{booktabs}
\usepackage{graphicx}
\usepackage{booktabs}
\usepackage{url}
\usepackage{acro}
\usepackage{xcolor}

\usepackage{geometry}
\usepackage{fancyhdr}

\fancypagestyle{firststyle}{
  \fancyhf{}
  \fancyhead[C]{2025 Asia Pacific Signal and Information Processing Association Annual Summit and Conference (APSIPA ASC)}
}

\DeclareAcronym{ITTS}{
  short = I$^2$TTS ,
  long  = Image-indicated Immersive Text-to-speech Synthesis
}
\DeclareAcronym{TTS}{
  short = TTS ,
  long  = Text-to-speech
}
\DeclareAcronym{CLIP}{
  short = CLIP ,
  long  = Contrastive Language-Image Pretraining
}
\DeclareAcronym{VITS}{
  short = VITS ,
  long  = Variational Inference with adversarial learning for end-to-end Text-to-Speech
}
\DeclareAcronym{SRC}{
  short = SRC ,
  long  = speech reverberation classifier
}
\DeclareAcronym{RIR}{
  short = RIR ,
  long  = room impulse responses
}
\DeclareAcronym{GeLU}{
  short = GeLU ,
  long  = Gaussian Error Linear Unit
}
\DeclareAcronym{CE}{
  short = CE ,
  long  = Cross-Entropy
}
\DeclareAcronym{MAS}{
  short = MAS ,
  long  = Monotonic Alignment Search
}
\DeclareAcronym{MCD}{
  short = MCD ,
  long  = mel cepstral distortion
}
\DeclareAcronym{ACC}{
  short = ACC ,
  long  = word accuracy
}
\DeclareAcronym{SRE}{
  short = SRE ,
  long  = Space Recognition Error
}
\DeclareAcronym{MFCCs}{
  short = MFCCs ,
  long  = mel-frequency cepstral coefficients
}
\DeclareAcronym{MOS}{
  short = MOS ,
  long  = Mean Opinion Score
}
\DeclareAcronym{SMOS}{
  short = SMOS ,
  long  = Similarity-Mean Opinion Score
}
\DeclareAcronym{IMOS}{
  short = IMOS ,
  long  = Immersive-Mean Opinion Score
}
\DeclareAcronym{NMOS}{
  short = NMOS ,
  long  = Naturalness-Mean Opinion Score
}
\DeclareAcronym{Grad-CAM}{
  short = Grad-CAM ,
  long  = Gradient-weighted Class Activation Mapping
}

\begin{document}

\title{I$^2$TTS: Image-indicated Immersive Text-to-speech Synthesis with Spatial Perception}

\author{
\authorblockN{
Jiawei Zhang\authorrefmark{1}\textsuperscript{1},
Tian-Hao Zhang\authorrefmark{1},
Jun Wang\authorrefmark{2},
Jiaran Gao\authorrefmark{1},
Ruijie Tao\authorrefmark{3},
Xinyuan Qian\authorrefmark{1}\textsuperscript{2},
and Xu-Cheng Yin\authorrefmark{1}
}

\authorblockA{
\authorrefmark{1} University of Science and Technology Beijing, Beijing, China \\
\authorrefmark{2} Tencent AI Lab, Shenzhen, China \\
\authorrefmark{3} National University of Singapore, Singapore
}
}

\maketitle
\thispagestyle{firststyle}
\pagestyle{fancy}

\footnotetext[1]{Work done during internship at Tencent AI Lab.}
\footnotetext[2]{Corresponding author}

\begin{abstract}
Controlling the spatial and stylistic characteristics of synthesized speech is essential for immersive and personalized applications such as virtual reality, gaming, and human-computer interaction. While recent \ac{TTS} systems have explored multi-modal conditioning, they often suffer from poor reverberation fidelity or degraded audio quality due to reliance on external vocoders. In this paper, we propose \ac{ITTS}, an end-to-end multi-modal \ac{TTS} framework that synthesizes high-quality, immersive speech from text and visual scene prompts. Our model leverages a CLIP-based image encoder with an adaptive adapter to extract scene-aware features, a Speech Reverberation Classifier (SRC) for refining acoustic-visual alignment during training, and a speaker encoder to enable zero-shot speaker generalization. Built upon a VITS backbone, \ac{ITTS} generates reverberant speech directly without requiring a separate vocoder. Experimental results demonstrate that our approach produces spatially accurate and natural-sounding speech, achieving superior performance in both subjective and objective evaluations. Project demo page: \url{https://spatialTTS.github.io/}
\end{abstract}

\section{Introduction}
\label{sec:intro}

Speech synthesis has seen remarkable advancements in recent years, driven by breakthroughs in deep learning and neural network architectures~\cite{wang2017tacotron,ren2019fastspeech,kim2021conditional,wang2023neural}. These innovations have enabled the generation of high-quality, natural-sounding speech across various applications, such as virtual assistants, accessibility technologies, and interactive media. Recently, numerous studies~\cite{casanova2022yourtts,guo2023prompttts,yang2024instructtts,guan2024mm} have explored the use of prompts to control specific speech characteristics, such as emotion, style, or speaker identity, enabling personalized and dynamic speech synthesis.

Reverberation is also a critical auditory feature that defines how sound interacts with its environment, profoundly influences the perception and intelligibility of speech~\cite{stan2002comparison}. While reverberation plays a vital role in creating immersive auditory experiences, most existing speech synthesis models are designed for neutral or non-reverberant environments, limiting their applicability in spatially aware or immersive contexts. To address this, recent efforts have begun integrating environmental context into the synthesis process. For instance, Environment-Aware TTS~\cite{tan2021environment} introduces an environment embedding extractor that learns environmental features from reference speech, improving the synthesis of speech tailored to specific acoustic settings by reducing intra-environment embedding distance and increasing inter-environment separation. Similarly, VoiceLDM~\cite{lee2024voiceldm} focus on mapping textual or audio descriptions into environmental feature vectors, effectively controlling the reverberation aspects of the generated audio. Furthermore, ViT-TTS~\cite{liu2023vit} presents a groundbreaking multi-modal \ac{TTS} task that generates speech with reverberation characteristics closely aligned with the acoustic properties of specific visual scenes. This innovation marks a significant advancement by integrating visual context into TTS. In addition, MS2KU-VTTS~\cite{he2025multi} explores immersive TTS by leveraging multi-source spatial knowledge, including image, depth, and semantic segmentation, to enhance spatial understanding. However, they depend on an external vocoder for waveform generation, which introduces synthesis artifacts, such as distortions and noise, and struggles to preserve fine-grained acoustic details like reverberation. Given that reverberation is inherently linked to environmental context, achieving robust and natural reverberation modeling necessitates a more integrated and end-to-end approach within the TTS pipeline.

To address this challenge, we have integrated scene-aware reverberation directly into the end-to-end speech synthesis process. To capture the acoustic characteristics of a given environment, we first use an image encoder to extract visual features from the scene's image. Specifically, we employ the \ac{CLIP}~\cite{radford2021learning} model for our image encoding process, as it effectively aligns visual and textual information, enabling the model to extract meaningful features representing the scene's environment. In addition to scene understanding, personalization is also crucial for immersive experience. We introduce a speaker encoder module that enables zero-shot speaker adaptation, allowing the model to synthesize speech with the voice characteristics of an unseen speaker based on a short reference speech. This module extracts speaker embeddings from reference speech and conditions the speech generation process accordingly, ensuring that the synthesized output reflects the target speaker identity. The entire pipeline is built on \ac{VITS}~\cite{kim2021conditional}, with modifications to incorporate the features that are relevant to the scene. Once immersive speech is generated, we further refine it using a \ac{SRC} model, which adjusts the reverberation and refines the reverberation characteristics to ensure accurate alignment with the prompt scene.

Our contributions are summarized as follows:
\begin{enumerate}
    \item We propose the first end-to-end multi-modal \ac{TTS} framework that synthesizes immersive and high-quality reverberant speech directly from text and visual scene prompts, built upon a VITS-based backbone without requiring a separately trained vocoder.
    \item We introduce a novel architecture that includes a CLIP-based image encoder to extract scene-aware visual representations and an SRC (Scene Relevance Classifier) module to refine the acoustic consistency between the synthesized speech and the visual environment, thereby improving scene precision. 
    \item Our system supports zero-shot speaker adaptation through a speaker encoder, enabling personalized and spatially grounded speech generation that realizes specific speakers in specific scenes.
    \item Extensive objective and subjective evaluations demonstrate that our method achieves state-of-the-art performance in generating spatially aligned, high-fidelity speech across diverse visual environments.
\end{enumerate}

\section{RELATED WORK}
\label{sec:related work}

\begin{figure*}[htbp]
  \centering
  \includegraphics[width=\textwidth]{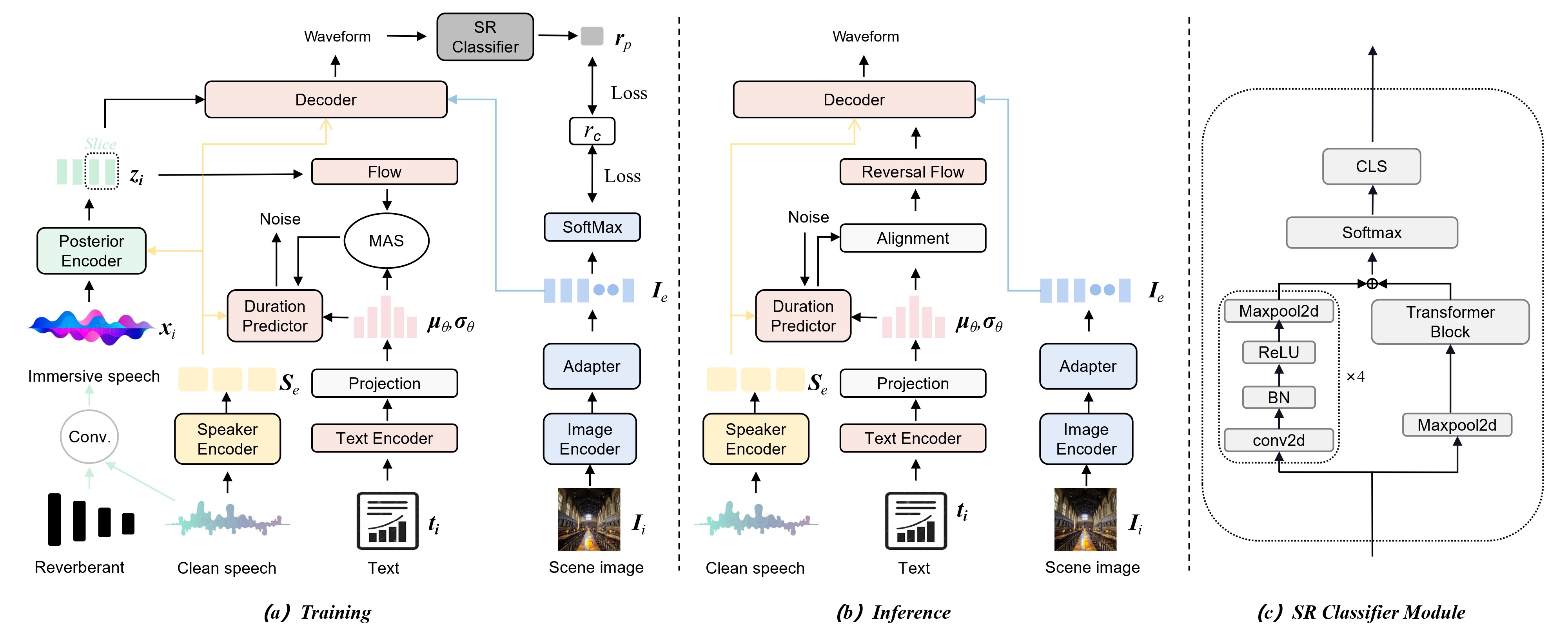}
  \caption{Block diagram of the \ac{ITTS} model ($\oplus$ indicates element-wise addition).}
  \label{fig:method}
\end{figure*}
\ac{TTS} has evolved significantly over the past few decades, moving from early rule-based systems~\cite{allen1987text,black1998festival,tokuda2002hmm} to modern deep learning-based approaches~\cite{wang2017tacotron}. This section reviews key developments in \ac{TTS}, with a focus on areas relevant to our proposed task: reverberation modeling, context-aware speech synthesis, and the integration of visual cues in speech generation.

\subsection{Standard Speech Synthesis}
Early speech synthesis systems, such as formant synthesis and concatenative synthesis, focused on generating speech by stringing together pre-recorded phonemes or sound units. Although these methods were able to produce intelligible speech, they often lacked naturalness and flexibility, particularly in varying acoustic environments. The advent of deep learning brought about a paradigm shift, with models like Tacotron~\cite{wang2017tacotron} that enables end-to-end TTS to produce highly natural and expressive speech. Since then, TTS has entered an explosive development with many models being explored, such as FastSpeech2~\cite{ren2020fastspeech}, Diff-TTS~\cite{jeong2021diff} and ProDiff~\cite{huang2022prodiff}. Recently, discrete token-based audio synthesis approaches, such as CLaM-TTS~\cite{kim2023clam}, VALL-E~\cite{wang2023neural}, AudioLM~\cite{borsos2023audiolm} and NaturalSpeech3~\cite{ju2024naturalspeech} formulate the generation process as a conditional language modeling task. Using extensive training data and the power of large language models, they significantly enhance speech quality and naturalness. However, these models typically generate speech in a neutral and non-reverberant environment, without considering the acoustic characteristics of the scene.

\subsection{Spatial Perception in Speech Synthesis}
Reverberation modeling has traditionally been addressed in the field of audio signal processing, where techniques such as convolution with room \ac{RIR} are used to simulate the effects of different acoustic environments on speech. In the context of speech synthesis, reverberation is often applied as a post-processing step, where the synthesized speech is convolved with an \ac{RIR} to achieve the desired effect~\cite{singh2021image2reverb, chen2022visual, somayazulu2024self}. Although effective, this approach treats reverberation as an integral part of the synthesis process rather than an afterthought. To address this limitation, recent research has started to integrate environmental cues directly into the synthesis pipeline. For instance, ViT-TTS~\cite{liu2023vit} introduces a multimodal speech synthesis paradigm where visual scenes are used to condition the generation of speech with appropriate reverberation. By leveraging visual prompts, it attempts to model the spatial context inherent in the image. Extending this direction, MS2KU-VTTS~\cite{he2025multi} takes a more holistic approach to immersive TTS by incorporating multi-source spatial information, including RGB images, depth maps, and semantic segmentations. However, due to their reliance on external vocoders, they may suffer from synthesis artifacts and limited control over fine-grained reverberation.

\subsection{Summary}
While speech synthesis has advanced significantly in producing natural and expressive outputs, the integration of environmental and spatial context remains a relatively nascent area. Post-processing methods for spatial effects, such as RIR convolution, have been standard but lack deep integration with the synthesis process. Recent efforts, have begun to explore multi-modal approaches that utilize scene prompts, yet challenges in speech quality and acoustic alignment persist due to their reliance on external vocoders and limited reverberation modeling fidelity.

Our approach addresses these limitations by tightly integrating scene-aware reverberation into the TTS pipeline. By treating environmental acoustics as an integral part of synthesis rather than a post-processing effect, our model achieves superior spatial alignment and speech quality, offering a more immersive and contextually accurate auditory experience.

\section{PROPOSED METHOD}
\label{sec:method}

The overall architecture of our proposed framework is shown in Fig~\ref{fig:method}. Our proposal is expected to generate speech that not only sounds natural, but also conveys a strong sense of spatial presence and speaker identity, guided by the scene and speaker prompts.

\subsection{Image Encoder}
In order to capture the acoustic characteristics of a given scene prompt, we first utilize the \ac{CLIP} model and an adapter to extract visual features from the scene image that represent the scene's likely acoustic properties.  The adapter module in our model is designed to efficiently integrate visual features of the scene into the speech synthesis pipeline. This module is simply comprised of a MLP network.

In practice, we first input the scene image $I_{i}$ into the \ac{CLIP}-based image encoder to extract the scene prompt features. After being fed to the adapter, we transform the scene prompt features into an acoustic space embedding $I_{e}$. To make this embedding close to the reverb category $r_{c}$, we guide it with a SoftMax layer and compute a \ac{CE} loss function with $r_{c}$.
Once obtaining the output embedding from the adapter module, we directly feed it into  the \ac{TTS} backbone. This integration allows the scene-aware information, encapsulated in the embedding, to influence various stages of the \ac{TTS} pipeline.

\subsection{Speaker Encoder}
To enable zero-shot speaker adaptation and ensure that the synthesized speech maintains consistent speaker identity, we introduce a dedicated speaker encoder module. This module extracts speaker embeddings from a short reference speech, allowing our model to imitate unseen speaker voices without retraining.

Concretely, inspired by YourTTS~\cite{casanova2022yourtts}, we adopt the StyleEncoder architecture from~\cite{min2021meta} to encode speaker-specific characteristics into a compact embedding $S_{e}$. This embedding is then projected via a linear layer and injected into the \ac{TTS} backbone through adaptive layer modulation, allowing the system to synthesize speech in the voice of an unseen speaker in a zero-shot manner.

\subsection{Speech Reverberation Classification}
To ensure that the reverberation characteristics of the synthesized speech are accurately matched to the visual scene, once the initial mel-spectrogram is generated, we further refine it using an \ac{SRC} model that classifies and adjusts the reverberation to ensure that it matches the scene. 

For the architecture illustrated in Fig.~\ref{fig:method} (b), the \ac{SRC} reveives the mel-spectrogram from the synthesis speech and predicts the class of reverberation. Specifically, the main network comprises four 2-D convolution blocks to capture local time-frequency features and a transformer~\cite{vaswani2017attention} block to obtain global dependencies across time frames. Then the features of the two blocks are concatenated to form a comprehensive feature representation fed into a softmax layer for further transformation, followed by a classification layer that outputs the final predictions. The predictions of the reverb class $r_{p}$ are finally computed by a \ac{CE} loss.

\subsection{\ac{TTS} Backbone}
We adopt \ac{VITS}~\cite{kim2021conditional} as the backbone of our system for its end-to-end design and high-quality output. It consists of a Posterior Encoder, Text Encoder, and Transformer-based flow for expressive latent modeling, with MAS and Duration Predictor for alignment and phoneme timing, and a Decoder to reconstruct the final waveform.

During training, the model receives text, immersive speech, speaker embedding, and scene embedding as input. The scene embedding, extracted from the CLIP-based image encoder and adapted via an MLP, guides the model to learn scene-aware reverberation characteristics. Moreover, speaker embedding $S_e$, obtained via a speaker encoder, allows the system to generalize to unseen speakers in a zero-shot fashion. Both embeddings are injected into the model through conditional modules, enabling joint learning of content, speaker identity, and spatial acoustics. Finally, the \ac{SRC} module verifies and refines the reverberation characteristics in the mel-spectrogram to better align with the input scene.

At inference time, the system takes input text, an image prompt, and an optional speaker reference. The adapted CLIP feature and speaker embedding modulate the synthesis process, allowing the decoder to generate immersive speech with appropriate reverberation and speaker identity. 

\section{EXPERIMENT}
\label{sec:experiment}
\begin{table*}[thb]
\centering
\caption{Experimental results for different models.}
\label{tab:results}
\renewcommand{\arraystretch}{1.2}
\setlength{\tabcolsep}{12pt}
\setlength{\textwidth}{16pt}
\scalebox{1.0}{
    \begin{tabular}{c|cccc|ccc}
        \hline
        ~&\multicolumn{4}{c|}{Objective metrics}&\multicolumn{3}{c}{Subjective metrics}\\
        \hline
        Model &WER(\%) $\downarrow$& MCD $\downarrow$ & SECS$\uparrow$& SRE(\%)  $\downarrow$ &NMOS $\uparrow$  & SMOS$\uparrow$ & IMOS $\uparrow$\\
        \hline
        GT & 4.7 & / &0.76&21.4&4.54&4.46&4.63\\
        Baseline&7.8&4.25&/&31.6&3.83&/&3.92\\
        ViT-TTS&8.6&4.58&/ &64.2&3.77&/&3.71\\
        MS2KU-VTTS&8.2&4.49&/&57.1&3.79&/&3.86\\
        \hline
        Proposed method w/o SR Classifier&8.0&4.29&0.53&39.7&3.82&4.16&3.81\\
        Proposed method w/o CLIP&7.8&4.27&0.57&30.8&3.85&4.19&3.83\\
        Proposed&7.6&4.22&0.62&27.2&3.98&4.21&3.96\\
    \end{tabular}}
\end{table*}

\subsection{Implementation Details}

For our experiments, we used three primary datasets: LJSpeech~\cite{ito2017lj} , VCTK~\cite{veaux2017cstr} and Image2Reverb~\cite{singh2021image2reverb}. These datasets provide the necessary diversity in speech, reverberated sound, and scene images to train and evaluate our proposed method. Specifically, we randomly convolve the clean speech data from the LJSpeech dataset and VCTK dataset with impulse responses from the Image2Reverb dataset. This process generates convolutional speech samples with reverberation effects that correspond to the visual scenes depicted in the images, which finally serves as the ground truth for training, providing a benchmark for evaluating the accuracy and naturalness of the synthesized output.

Additionally, we pretrained the \ac{SRC} model on the convolutional speech samples over 300 epochs. Once integrated into the \ac{TTS} framework, we froze its parameters and only used the model to calculate the classification loss of the generated speech. The \ac{TTS} framework was trained for 200K iterations using AdamW optimizer on 3 NVIDIA GeForce RTX 3090 GPUs.

For the comparative analysis, we conducted experiments on the following systems: 1) GT: the ground-truth convolutional speech. 2) Baseline: the \ac{VITS} backbone trained on the clean speech followed by convolution with RIR. 3) ViT-TTS. 4) MS2KU-VTTS. 5) Proposed method w/o SR Classifier: our proposed framework that without \ac{SRC} module. 6) Proposed method w/o CLIP: Our proposed framework utilizing a convolution network replaces the CLIP encoder. 

\subsection{Evaluation Metrics}
To assess the effectiveness of our proposed method, we first evaluate the quality of the generated speech and the correspondence between the scene and the reverberation using some objective metrics including Word Error Rate (WER), \ac{MCD} ~\cite{kubichek1993mel}, Speaker Encoder Cosine Similarity (SECS)~\cite{casanova2021sc} and \ac{SRE}. WER measures the accuracy of the synthesized speech in terms of word recognition. \ac{MCD} measures the spectral distance between the synthesized speech and the ground truth speech. SECS is used to assess the speaker similarity between the input and generated speech. \ac{SRE} measures the correctness of the room acoustics of the generated speech. 

We also conduct subjective evaluations to assess perceptual aspects that are difficult to capture quantitatively. We use a \ac{MOS}~\cite{sisman2020overview} with 95 $\%$ confidence intervals to assess speech quality, where listeners are asked to rate the audio on a scale from 1 to 5. We employ the \ac{NMOS} to gauge the naturalness of synthesized speech to human listeners, evaluating its overall fluidity, expressiveness, and resemblance to human speech. Simultaneously, we utilize the \ac{SMOS} to measure the perceived similarity between the speaker's voice in the input and the generated speech.  Furthermore, We test \ac{IMOS} for matching between scene prompt and spatial characteristics of speech to evaluate the effectiveness of our model in aligning with human perceptions of the scene's acoustics. We engaged 20 listeners each evaluating clarity, reverberation accuracy, and naturalness on 6 samples of test set. Standard deviations across these scores were small (\textless0.25), indicating consistent evaluations.

\subsection{Experimental Results}
Table~\ref{tab:results} presents the objective and subjective evaluation results across different models. Compared to the baseline and prior scene-aware TTS systems (ViT-TTS and MS2KU-VTTS), our proposed method consistently achieves superior performance across nearly all metrics.From the objective perspective, our full model attains the lowest WER (7.6\%) and MCD (4.22), indicating enhanced intelligibility and spectral fidelity. Furthermore, it shows a notable improvement in SRE, reducing the mismatch between predicted and target scene acoustics to 27.2\%, a significant gain over ViT-TTS (64.2\%) and MS2KU-VTTS (57.1\%). The SECS also improves to 0.62, demonstrating better preservation of speaker identity.In terms of subjective quality, our model achieves the highest scores on NMOS (3.98), SMOS (4.21), and IMOS (3.96), reflecting improved naturalness, speaker similarity, and perceived scene-spatial alignment. Notably, even ablation variants (w/o CLIP or SR Classifier) outperform existing methods, highlighting the robustness of our architecture.These results confirm the effectiveness of integrating both visual scene prompts and speaker identity encoding into an end-to-end TTS framework, enabling the generation of immersive, high-quality, and scene-consistent reverberant speech.

\subsection{Environments analysis}
\begin{table}[tb]
\centering
\caption{Experimental results for different environment.}
\label{tab:environments}
\renewcommand{\arraystretch}{1.3}
\setlength{\tabcolsep}{12pt}
\setlength{\textwidth}{16pt}
\scalebox{.85}{
    \begin{tabular}{c|cc|cc}
        \hline
        ~&\multicolumn{2}{c|}{Wide environment}&\multicolumn{2}{c}{Narrow environment}\\
        \hline
        Model & SRE(\%)  $\downarrow$  & IMOS $\uparrow$ & SRE(\%)  $\downarrow$  & IMOS $\uparrow$\\
        ViT-TTS&58.2&3.73&67.1&3.68\\
        MS2KU-VTTS&54.7&3.89&62.8&3.81\\
        Proposed&25.3&4.01&28.6&3.90\\
        
    \end{tabular}}
\end{table}
To further evaluate our model’s robustness across diverse acoustic conditions, we analyze performance separately in wide and narrow environments. As shown in Table~\ref{tab:environments}, our proposed method significantly outperforms ViT-TTS and MS2KU-VTTS in both scenarios. In wide environments, which typically exhibit longer reverberation tails and more diffuse reflections, our model achieves the lowest Scene Reverb Error (SRE) at 25.3\% and the highest IMOS score of 4.01. This suggests that the model is more capable of capturing complex spatial cues from wide scenes, leading to more immersive and perceptually accurate reverberation. In narrow environments, which feature shorter reverberation times and tighter spatial constraints, our method again yields superior results with an SRE of 28.6\% and an IMOS of 3.90. Compared to the SREs of 67.1\% and 62.8\% from ViT-TTS and MS2KU-VTTS respectively, our approach demonstrates better adaptability and scene-awareness. These findings confirm that the integration of visual scene prompts and the refinement by the SRC module enable our model to generalize well across spatial contexts. The consistently higher IMOS scores also reflect that human listeners perceive the reverberation in our outputs to be more faithful to the corresponding visual scenes, regardless of environmental scale.

\subsection{Visualization}
\begin{figure}[tb]
  \centering
  \includegraphics[width=\columnwidth]{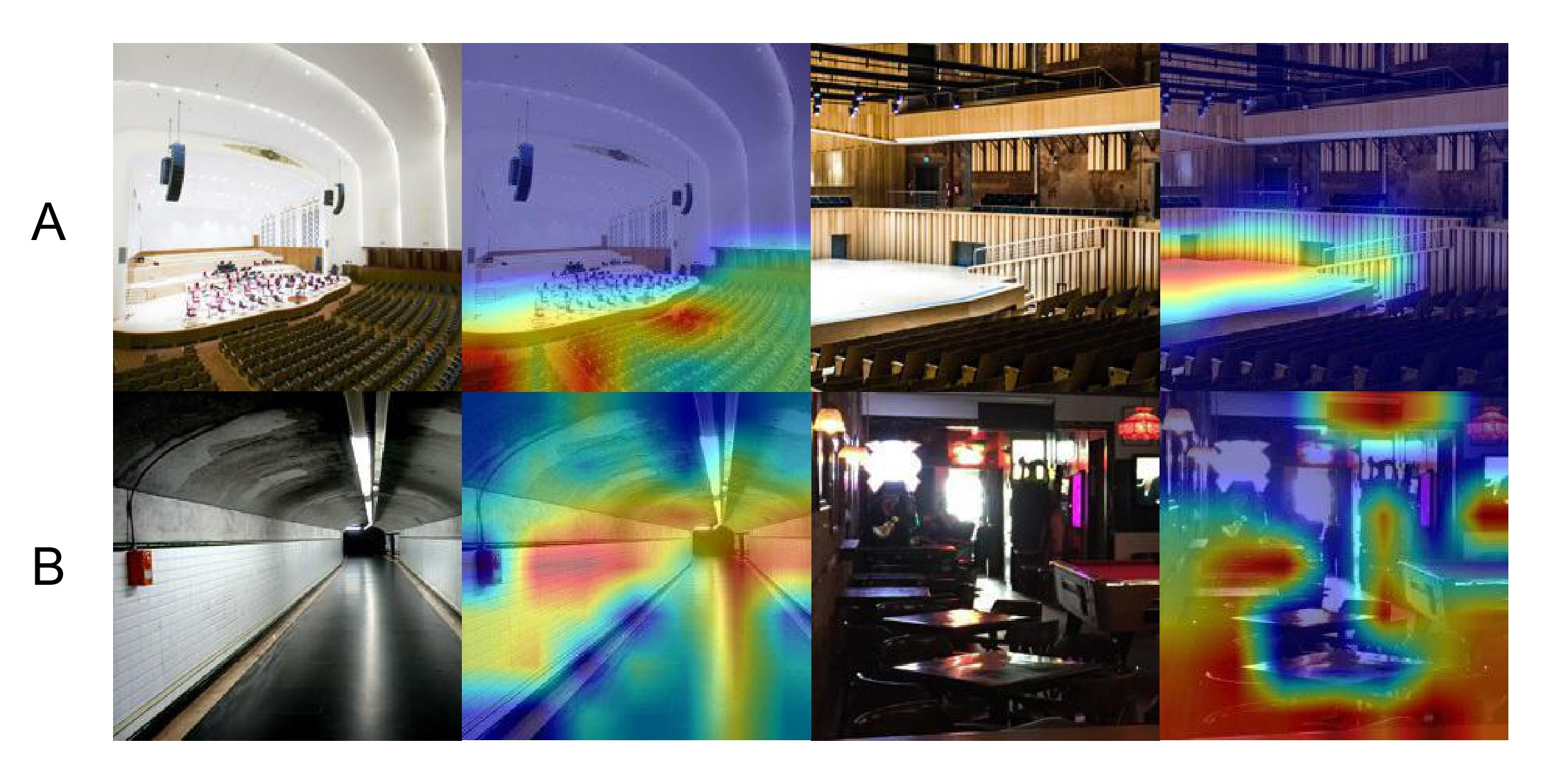}
  \caption{\ac{Grad-CAM}s for images passed the scene prompt encoder, showing movement towards more textured areas for (A) a wide, and (B) a narrow environment.}
  \label{fig:grad-cam} 
  \vspace{-3mm}
\end{figure} 

To gain insight into which visual features are most influential in our encoder, we employ \ac{Grad-CAM}~\cite{selvaraju2017grad}, a widely used technique applied to visually interpret the image regions that contribute the most to scene understanding. \ac{Grad-CAM} allows us to visualize the areas of the input image that contribute most significantly to the scene-aware feature extraction process. We produce such maps for our test images with the scene prompt encoder as shown in Fig.~\ref{fig:grad-cam}. The original images are shown on the left, with the corresponding \ac{Grad-CAM} heatmaps on the right. These heat maps highlight areas the model considers crucial for extracting image features relevant to the acoustic environment. The color scale, which ranges from blue to red, indicates increasing levels of importance. For example, in the wide environment (A), the highlighted regions emphasize broad structural elements such as walls and seating arrangements. In contrast, in the narrow environment (B), attention is focused on confined linear features such as hallways. Experimental results demonstrate that the detected visual objects aid in \ac{TTS} systems with reverberation.

\section{Conclusion}
\label{sec:conclusion}

In this work, we presented a novel approach to end-to-end multi-modal \ac{TTS} system that incorporates visual scene prompts to guide the generation of immersive, contextually appropriate speech. Our method extends the capabilities of traditional \ac{TTS} by introducing a framework that integrates spatial and scene-aware acoustic features into the synthesis process, addressing limitations in existing models. We demonstrated that our model achieves high-quality scene and reverb matching without deteriorating the naturalness of the speech. The combination of objective metrics and subjective evaluations shows that our model successfully balances speech quality with scene awareness, paving the way for more immersive and context-sensitive applications in virtual environments, interactive media, and other fields requiring spatially-aware audio.

\section{Acknowledgement}
\label{sec:acknowledgement}
This work is supported by the National Natural Science Foundation of China under Grant No. 62306029, the Beijing Natural Science Foundation under Grants L233032,  Shenzhen Research Institute of Big Data under Grant No. K00120240007 and CCF-Tencent Rhino-Bird Fund.







\bibliographystyle{IEEEtran}
\bibliography{main}
\end{document}